# The Optical Properties and Plasmonics of Anisotropic 2-Dimensional Materials

*Chong Wang, Guowei Zhang\*, Shenyang Huang, Yuangang Xie, and Hugen Yan\**

Dr. C. Wang, S. Y. Huang, Y. G. Xie, Prof. H. Yan
State Key Laboratory of Surface Physics
Key Laboratory of Micro and Nano Photonic Structures (MOE)
and Department of Physics
Fudan University
Shanghai 200433, China
E-mail: hgyan@fudan.edu.cn

Prof. G. W. Zhang
MIIT Key Laboratory of Flexible Electronics (KLoFE)
Xi'an Institute of Flexible Electronics
Northwestern Polytechnical University
Xi'an 710072, Shaanxi, China
E-mail: iamgwzhang@nwpu.edu.cn

**Abstract:**

In the fast growing two-dimensional (2D) materials family, anisotropic 2D materials, with their intrinsic in-plane anisotropy, exhibit a great potential in optoelectronics. One such typical material is black phosphorus (BP), with a layer-dependent and highly tunable band gap. Such intrinsic anisotropy adds a new degree of freedom to the excitation, detection and control of light. Particularly, hyperbolic plasmons with hyperbolic $q$-space dispersion are predicted to exist in BP films, where highly directional propagating polaritons with divergent densities of states are hosted. Combined with a tunable electronic structure, such natural hyperbolic surfaces may enable a series of exotic applications in nanophotonics. In this review, the anisotropic optical properties and plasmons (especially hyperbolic plasmons) of BP are discussed. In addition, other possible 2D material candidates (especially anisotropic layered semimetals) for hyperbolic plasmons are examined. This review may stimulate further research interest in anisotropic 2D materials and fully unleash their potential in flatland photonics.

# 1. Introduction

The past decade has witnessed significant progress in the field of two-dimensional (2D) materials. Strong interactions with light have been demonstrated in atomically thin 2D materials.[1-12] For example, single-layer graphene can absorb 2.3% of the incident light in the visible to near-IR spectral range, which is associated with interband transitions.[1, 2] More surprisingly, for intraband transitions in the terahertz range, the light extinction for doped single-layer graphene can even reach 60%.[4] Such strong optical absorption in 2D materials provides promising platforms for future optoelectronic applications, such as photodetectors,[13-16] light emitters[17, 18] and optical modulators.[19, 20]

Beyond graphene, transition metal dichalcogenides (TMDCs) (e.g., $MoS_2$, $MoSe_2$, $WS_2$, $WSe_2$) are also important members of the 2D materials family. They possess many intriguing properties, such as spin-valley coupling,[21-24] strong many-body interactions,[7-9] and indirect-to-direct bandgap transition in the monolayer limit.[25, 26] Their optical absorption is mainly in the visible spectral range, and inherently isotropic (identical to that of graphene). In addition, hexagonal boron nitride (hBN), with an insulating bandgap of ~6 eV, is another well-known 2D material that is widely used as an excellent dielectric in 2D material-based electronics.[27] Although anisotropic light emission from monolayer hBN has been demonstrated in the visible range, this anisotropy is not intrinsic and is instead due to vacancy-related defects.[28]

Recently, anisotropic 2D materials with intrinsic in-plane band anisotropy,

summarized by Li et al.,[29] have attracted significant attention. Among them are black phosphorus (BP), ReX$_2$ (X=S, Se), MX (M=Sn, Ge; X=S, Se), etc., with reduced crystal symmetry. Compared to graphene and TMDCs, one of the landmark features for anisotropic materials is their capability to detect and control light polarization. As an elemental 2D material beyond graphene, BP has been widely studied due to its tunable direct bandgap and relatively high mobility. In 2014, Li et al. reported the pioneering work on 2D BP transistors, with the carrier mobility reaching ~1000 cm$^2$/(V·s) at room temperature, showing very encouraging device performance.[30] BP has a direct bandgap that exhibits a strong layer dependence, ranging from 0.3 eV (bulk) to 1.7 eV (monolayer).[31-34] It covers a broad frequency range from visible to mid-IR, filling up the gap between the most popular graphene and TMDCs. Moreover, the bandgap can be further tuned by strain[31, 35-40] and electric fields,[41-44] and even fully closed with a semiconductor-to-metal transition.[37, 41] The sizeable and highly tunable direct bandgap, combined with the relatively high carrier mobility, makes BP a promising candidate in IR photonics and optoelectronics.

Most importantly, BP exhibits a unique in-plane anisotropy, originating from the puckered atomic structure. Inside each BP layer, which consists of two sublayers, P atoms are covalently connected to three neighboring atoms to form the puckered hexagonal unit, while the BP layers are held together by weak van der Waals (vdWs) interactions. The low lattice symmetry of BP leads to two distinct in-plane directions: armchair (AC, perpendicular to the pucker) and zigzag (ZZ, along the pucker), as illustrated in **Figure 1**a. Previous studies have demonstrated strong anisotropy in electrical,[45-47] optical,[31, 32, 46, 48-52] thermal[53-55] and mechanical[56, 57] properties, distinguishing BP from graphene and TMDCs.

Most attention has been given to the interband transitions of BP in terms of its optical properties. Nevertheless, the plasmonic response from free carriers (intraband transitions) in this intrinsically anisotropic material has also been deemed interesting. Plasmons, the collective excitation of charge carriers, have been extensively studied in noble metals[58] and conventional two-dimensional electron gases (2DEGs).[59] In the past few years, tremendous attention has been given to graphene plasmons, with a remarkable manifestation of high tunability and low loss.[3, 4, 60-73] Recently, a number of theoretical efforts have been devoted to studying the plasmonic properties of BP, where anisotropic plasmon dispersions occur due to its band anisotropy.[74-95] Fascinatingly, the hyperbolic plasmon polaritons, whose iso-frequency contour is a hyperbola, are predicted to naturally exist in BP films, which originate from the coupling between the intraband and interband optical conductivity.[77, 79, 95] This special plasmon topology leads to an anomalously large photonic density of states and directional propagating plasmonic rays, which promises intriguing applications for planar photonics.[96-99] Moreover, the tunability of the band structure by electrical gating and strain in BP[27, 100] and the possibility of forming heterostructures with other 2D materials[101-103] make the plasmon in BP exhibit several unique potential applications that cannot be realized in metasurfaces composed of artificial structures.[104, 105]

Although hyperbolic plasmons have not yet been demonstrated in BP, experimental studies have reported the signature of anisotropic plasmon modes by near-field and far-field methods.[106-109]

In future device applications, large-area and high-quality thin-layer BP is highly desirable. Although the synthesis of BP is not the focus of this review, it is necessary to briefly introduce the process, since it lays the foundation for BP applications. Currently, high-quality samples are primarily obtained by mechanical exfoliation, with the size of typically several to tens of μm and an apparent drawback of a low yield. Hanlon et al. adopted an alternative method of liquid phase exfoliation.[110] The produced flakes are typically ~1 μm in size and flakes with different thicknesses are usually mixed together. This severely limits the practical applications of BP. For quite some time efforts have been devoted to the synthesis of thin-layer BP. Xia group made progress in 2015, reporting an approach to synthesize millimeter-sized thick BP films (tens of nm in thickness), but with a low sample quality.[111] Later, the same group largely improved the quality of synthesized BP films, with a carrier mobility up to 160 $cm^2/(V·s)$ at room temperature.[112] So far, it is still quite challenging to directly synthesize thin-layer BP with a large area and high quality, requiring further efforts and is long-waiting for the applications of BP.

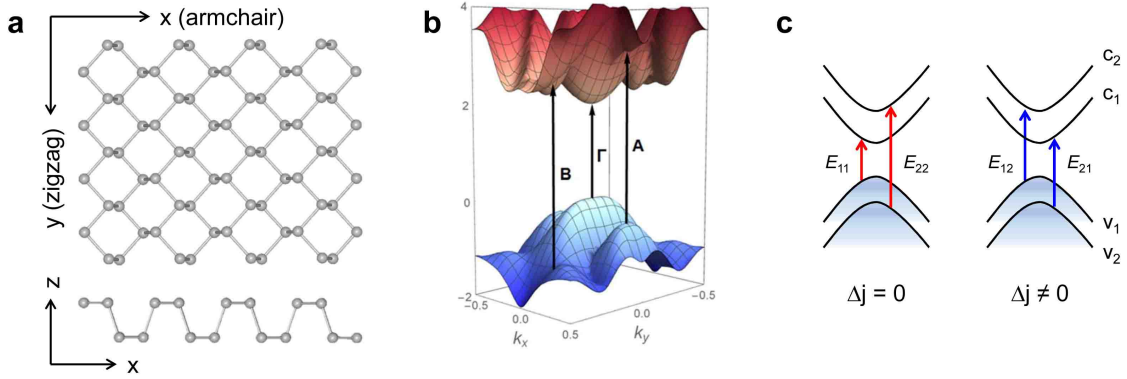

**Figure 1.** Lattice structure and band structure of a monolayer BP. a) Top (upper panel) and side (lower panel) views of the puckered lattice structure of monolayer BP, with two distinct in-plane directions: AC and ZZ. b) The 2D band structure of a monolayer BP. The significant optical transitions associated with the Van-Hove singularities (points Γ, A and B) are indicated by arrows. Reproduced with permission.[113] Copyright 2015, IOP Publishing. c) Schematic illustrations of optical transitions from valence subbands to conduction subbands. The red and blue arrows indicate the main optical transitions ($\Delta j = 0$) and hybrid optical transitions ($\Delta j \neq 0$), respectively.

In this review, we will take BP as a representative example of anisotropic 2D materials for the following reasons. (i) Direct bandgap. BP always has a direct bandgap, while most other anisotropic 2D materials, such as $ReSe_2$, SnS, etc., have an indirect bandgap,[29] inherently limiting their applications in optoelectronics. (ii) Large tunability. The bandgap of BP can be tuned to a wide range (0.3-1.7 eV, or mid-IR to visible range) simply by changing the layer number due to strong interlayer interactions. Moreover, the tuning range can be further extended by the strain and electrical methods. (iii) IR applications. While most of the semiconducting 2D materials (including anisotropic materials) have a large (or visible) bandgap, BP is one of the few that can reach the mid-IR range. Thus, many important IR applications are available for BP, such as optical communications and thermal imaging. In addition, the moderate bandgap enables coupling between intraband and interband transitions, which is vital for the realization of hyperbolic plasmons. (iv) High carrier mobility. BP exhibits a relatively high carrier mobility. A recent study by Long et al. reported a significantly improved mobility of up to 5200 cm$^2$/(V·s) at room temperature in high-quality few-layer BP samples encapsulated by hBN. The mobility can be further increased to 45000 cm$^2$/(V·s) at cryogenic temperatures,[114] making BP another layered material beyond graphene that exhibits the quantum Hall effect.[115] Hence, we only focus on BP here in order to discuss the optical and plasmonic properties with a special emphasis on the anisotropic character. This paper is structured as follows: (1) We start with a brief introduction. (2) This is followed by a discussion of anisotropic optical properties and polarized photodetection in BP. (3) Then we discuss the plasmonic properties of anisotropic 2D materials in BP, followed by a perspective on anisotropic layered semimetals as another choice for the realization of natural hyperbolic plasmons.

## 2. Optical Properties of Anisotropic 2D Materials

### 2.1 Polarized Absorption

Band parameters are crucial to semiconductors, since they largely determine the effective mass, carrier mobility and bandgap. Thus, the band structure is closely related to the transport and optical properties of semiconductors. An early experimental work for bulk BP dates back to 1953, with the bandgap determined to be 0.33 eV from electrical conductivity measurements.[116] Soon after the pioneering work on 2D BP transistors in 2014,[30] the band structure of BP was theoretically examined with the thickness ranging from bulk to monolayer.[33, 34] The theory shows that bulk BP is a direct-gap semiconductor with a gap size of ~ 0.3 eV, which is consistent with previous experimental values. The conduction band minimum (CBM) and valence band maximum (VBM) are both located at the Z point of the 3D Brillouin zone, while for the mono- and few-layer cases, the CBM and VBM are both shifted to point Γ of the 2D Brillouin zone, and the bandgap increases with a decreasing thickness. Most interestingly, the theory predicts a highly anisotropic band dispersion in BP, with the conduction and valence bands highly dispersive in the AC direction and nearly flat in the perpendicular ZZ direction.[53]

Low et al. reported the first theoretical results for the optical conductivity of BP films (thickness > 4 nm or 8L) using the Kubo formula.[52] With this film thickness, the excitonic effect can be excluded due to significant dielectric screening. The authors found that the optical absorption in BP is highly dependent on the thickness, doping and light polarization. They also predicted multiple optical transitions between the quantized subbands $E_{c,v}^j$ ($j$ is the subband index) in the absorption spectra due to the quantum confinement normal to the plane, akin to quantum wells (QWs). In symmetric QWs, only optical transitions with $\Delta j = 0$ are allowed for normal light incidence. Nevertheless, vertical electric fields across the BP layers can relax this selection rule, leading to hybrid optical transitions between conduction and valence subbands with different subband indices, as shown in Figure 1c.[117] Additional absorption features ($\Delta j \neq 0$) between the main peaks ($\Delta j = 0$) were experimentally observed in few-layer BP, possibly caused by unintentional doping, which breaks the symmetry of BP QWs.[31] Via controllable electrical doping, the intensity of hybrid peaks is expected to be tuned, along with the transfer of oscillator strength from the main peaks. In addition, the theoretical results by Low et al. show strong anisotropy for optical absorption in BP. This is due to the mirror symmetry in the *x-z* plane. As a consequence, optical transitions for light polarized along the ZZ direction are strictly forbidden.

In detail, Li et al. explained optical anisotropy based on a symmetry analysis.[118] Tran et al. then proposed a more material-centered way to understand this anisotropy by checking the dipole transition matrix.[113] Briefly, in the single-particle picture, the optical absorption coefficient can be expressed as follows:

$$\alpha(\omega) = \frac{1}{\omega} \sum_{vc\vec{k}} \left| \langle v\vec{k} | \vec{e} \cdot \vec{r} | c\vec{k} \rangle \right|^2 \delta(\hbar\omega - E_{c\vec{k}} + E_{v\vec{k}}) \tag{1}$$

where $\langle v\vec{k} | \vec{e} \cdot \vec{r} | c\vec{k} \rangle$ is the dipole transition matrix and the δ-function conserves the energy and is proportional to the joint density of states (JDOS), which fully determine the optical absorption strength (or so-called oscillator strength). Usually, optical resonance occurs at the Van-Hove singularity (VHS) in the band structure. Presented in Figure 1b is the calculated 2D band structure of a monolayer BP, in which there are three points labeled as Γ, A and B. At point Γ, the transition energy from the valence band to conduction band is 2.0 eV, which corresponds to the absorption edge for the AC polarization, while at points A and B, the transition energies are both approximately 3.8 eV, corresponding to the absorption edge for the ZZ polarization. By carefully checking the dipole transition matrices at these three specific points, they found that at point Γ the value is zero (nonzero) for ZZ (AC) polarization, which means that the optical transition is strictly forbidden for ZZ polarization while allowed for AC polarization. This can well explain the observed optical anisotropy in BP. Point A is similar to the case of point Γ, i.e., optical transition is only allowed for AC polarization. The only difference is that the transition energy is changed to 3.8 eV. At point B, the optical transitions for AC and ZZ polarizations are both allowed, without apparent optical anisotropy.

Experimentally, a few groups have demonstrated the optical anisotropy of BP in the visible and mid-IR range, both for few-layer and bulk cases.[31, 32, 46, 119] Xia et al. performed IR extinction ($1-T/T_0$) measurements for thick BP films (thickness: ~30 nm) on Si/SiO$_2$ substrates.[46] The extinction spectrum indicates an absorption edge at approximately 2400 cm$^{-1}$ (~0.3 eV), which is in good agreement with the previously reported bulk value. In addition, they showed that the extinction at approximately 2400 cm$^{-1}$ periodically changes with the light polarization angle, demonstrating strong optical anisotropy. Due to the severe interference caused by the substrate, the extinction spectra are thus largely modified and deviated from the intrinsic optical conductivity of BP.[52]

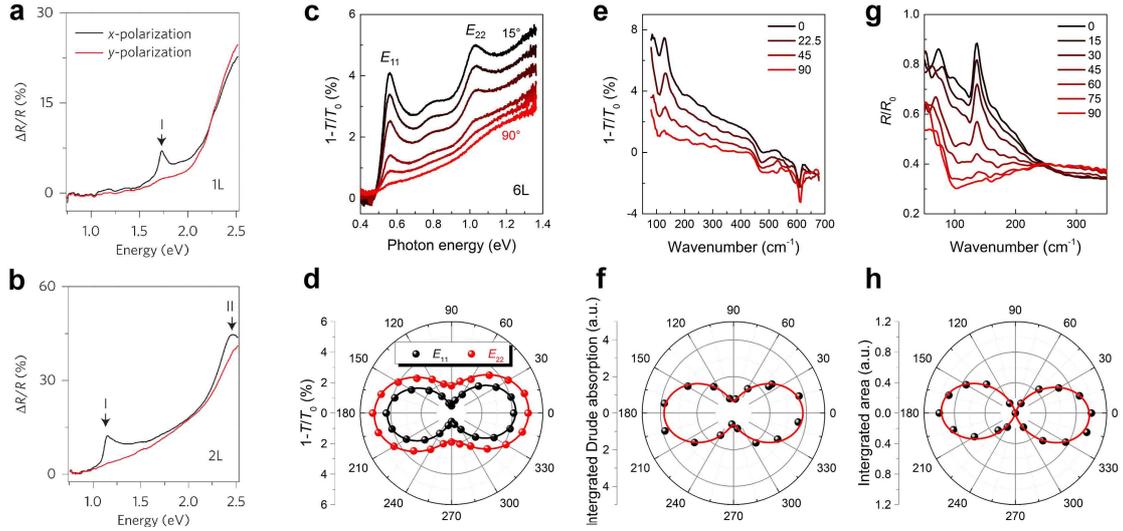

**Figure 2.** Polarized absorption in BP. Reflectance ($\Delta R/R_0$) spectra of a) 1L and b) 2L BP in the visible to near-IR range under AC ($x$) and ZZ ($y$) polarizations, respectively. Peaks I and II denote the first (bandgap) and second subband transitions, respectively. Reproduced with permission.[32] Copyright 2017, Springer Nature. c) Extinction ($1-T/T_0$) spectra of a 6L BP, with different light polarizations in the near- to mid-IR range. The labels $E_{11}$ and $E_{22}$ denote optical transitions between the first and second pair of subbands, respectively, which are sketched in the left panel of Figure 1c. d) Extinction at peaks $E_{11}$ and $E_{22}$, as a function of the polarization angle $\theta$. The solid lines are $\cos^2\theta$ fits. Reproduced with permission.[31] Copyright 2017, Springer Nature. e) Extinction spectra of thick BP (thickness > 100 nm) with different light polarizations, clearly showing an anisotropic Drude response due to intraband transitions. f) Polarization dependence of the integrated Drude absorption from 80 to 400 cm$^{-1}$, extracted from the spectrum fitting by a typical Drude-Lorentz model. The solid line is a $\cos^2\theta$ fit. g) Reflectance ($R/R_0$) spectra of thick BP (another sample different from that in Figure 2e) in the frequency range of 50-350 cm$^{-1}$ with different light polarizations, where an Au mirror is used as the reference. A prominent peak appears at ~136 cm$^{-1}$, due to IR phonon ($B_{1u}$ mode) absorption. h) Polarization dependence of the integrated area of the phonon absorption peak, normalized to the maximum value. The solid line is a $\cos^2\theta$ fit. 0° and 90° respectively denote the AC and ZZ polarizations for all the three samples in Figure 2c-2h.

Li et al. performed reflectance ($\Delta R/R_0$) measurements for 1-5L BP on sapphire substrates in the visible to near-IR range (0.75-2.5 eV).[32] The spectra show prominent absorption peaks for AC polarized light, with the peak positions markedly shifted with the layer number. For ZZ polarized light, no absorption resonances were observed (**Figure 2**a), demonstrating strong optical anisotropy in 1-5L BP. Both for AC and ZZ polarizations, the spectra are complicated by a steep and broad background, presumably due to nonresonant contributions. In addition to the lowest bandgap transitions, higher-order resonances were also observed (Figure 2b), originating from strong interlayer interactions and *z*-direction quantum confinement in the few-layer BP.

Zhang et al. reported the first systematic IR study of few-layer BP via Fourier transform infrared (FTIR) spectroscopy, with a thickness of 2-15L and photon energies of 0.25-1.36 eV.[31] Multiple absorption peaks have been observed, arising from optical transitions between quantized subbands. This uncovers the evolution of the band structure, with a strong thickness and polarization dependence. For BP samples of each thickness, the IR spectrum is unique and can be readily served as its fingerprint. Figure 2c presents the typical IR extinction spectra of a 6L BP under different light polarizations. Two prominent peaks (labeled as $E_{11}$ and $E_{22}$) can be identified, which are assigned to excitonic resonances related to transitions between the first ($v_1 \rightarrow c_1$) and second ($v_2 \rightarrow c_2$) pairs of subbands, respectively, as sketched in Figure 1c. For thicker BP films (e.g., 13L and 15L), at least four peaks appear within the measurement range. As summarized in Figure 2d, the two peaks of the 6L sample both show a strong polarization dependence, with the strongest absorption for AC polarization and much less absorption for ZZ polarization. This large optical anisotropy applies to all measured few-layer BP samples, making IR spectroscopy an easy and accurate method for crystal orientation determination. This technique is obviously superior to polarized Raman spectroscopy, since both the excitation wavelength and sample thickness complicate the polarization behavior for the latter.[120] Additionally, Jiang et al. provided a more visualized view of the optical anisotropy of BP via scanning polarization modulation microscopy (SPMM).[121] Other than the optical techniques, the strong anisotropy of BP has also been demonstrated by electron energy-loss spectroscopy (EELS).[122]

The anisotropic band structure of BP can also be manifested by the absorption of light from free carriers, i.e., the Drude response. Since the interband transition energy lies in the visible to mid-IR range for different thicknesses, the conductivity of BP in the far-IR range is dominated by the intraband excitations (Drude component). Figure 2e shows the extinction spectra of thick BP (thickness > 100 nm) with different light polarizations from the transmission measurements. A large Drude response can be found for AC polarization due to the smaller effective mass, since the Drude weight of normal carriers can be expressed as $D_k = \pi n e^2 / m_k$,[123] where $k$ denotes the crystal orientation (AC and ZZ), $m_k$ is the effective mass along the $k$ axis, and $n$ is the carrier density. In addition to the Drude absorption, an IR active phonon mode at ~ 136 cm$^{-1}$ can be observed in Figure 2e, which shows an anisotropic response as well. In the 1980s, optical absorption from this IR active phonon in bulk BP was reported, which has been attributed to the B$_{1u}$ phonon mode.[124-126] The polarization-dependent phonon absorption can also be observed in the reflectance spectra (Figure 2g). Figure 2f and 2h show the integrated areas of the Drude response and phonon absorption peak fitted by a typical Drude-Lorentz model.[105] A larger Drude weight and phonon absorption can be found for AC polarization, and the effective mass ratio can be derived from the fitted

Drude weights with $\frac{m_{zz}}{m_{AC}} = \frac{D_{AC}}{D_{ZZ}} \approx 5$, confirming the highly anisotropic band structure of BP.

**2.2 Polarized Photoluminescence**

Previous theory predicts that BP always possesses a direct bandgap, regardless of its thickness. Strong photoluminescence (PL) has indeed been observed in BP from monolayer to bulk by several groups,[31, 32, 45, 48, 49, 127-129] verifying their direct-gap character. An early work by Zhang et al. reported the observation of layer-dependent PL in 2-5L BP in the visible to near-IR range.[49] Different from optical absorption spectroscopy, PL spectroscopy is very sensitive to impurity and defects. Given that monolayer BP degrades very fast in air, defects are very easy to introduce. Therefore, in some cases the observed optical emission is from defects instead of the intrinsic excitonic emission. This leads to the confusing results previously reported by different groups, with the PL position ranging from 1.30 to 1.73 eV for monolayer BP.[31, 32, 48, 127] The PL emission from defects was also observed in both monolayer[130] and bilayer[131] BP at 1.36 eV and 1.0 eV, respectively, which is much lower than those for the intrinsic bandgap emission. This defect nature was also confirmed by a sublinear relation of the emission intensity versus the excitation intensity. Nevertheless, defect engineering is highly desirable for realizing single photon emission in BP, so it merits further efforts.

Wang et al. reported the observation of polarized PL in monolayer BP, revealing highly anisotropic excitons.[48] **Figure 3**a presents the polarization-resolved PL spectra of a monolayer BP at room temperature, with the excitation and detection light polarized in either the *x* (AC) or *y* (ZZ) directions, respectively. The PL spectrum shows a broad linewidth of ~150 meV, possibly due to the poor sample quality. Figure 3b shows that the PL intensity is strongly dependent on the excitation polarization, which is a direct consequence of the anisotropic optical absorption in BP. Interestingly, the PL emission also shows strong anisotropy with respect to the detection polarization, indicating that the emitted light is always linearly polarized in the AC direction.

Li et al. observed strong PL emission in 1-3L BP protected by thin hBN flakes, with the emission peak matching the bandgap absorption (Figure 3c-3e).[32] As expected, the PL intensity shows a strong polarization dependence that is strongest along the AC direction and absent along the ZZ direction. Note that the Stokes shift, defined as the energy difference between the PL and absorption peaks, is as large as 20-40 meV, which could arise from phonon sidebands, photoinduced structural relaxation and/or defects and impurities. In high quality samples, the Stokes shift is expected to diminish.

Complementing the visible to near-IR PL studies in thin-layer BP (thickness < 5L), Chen et al. reported the first mid-IR PL emission from thicker BP (thickness > 9L) using an FTIR spectrometer combined with an IR microscope.[129] The PL spectra show strong dependences on the sample thickness and temperature, with the peak position well described by a tight-binding model, as in the absorption case. Moreover, they showed that the mid-IR PL intensity of BP is comparable to that of traditional InAs QWs, opening up avenues for potential applications of tunable (through thickness) mid-IR light emitting and lasing.

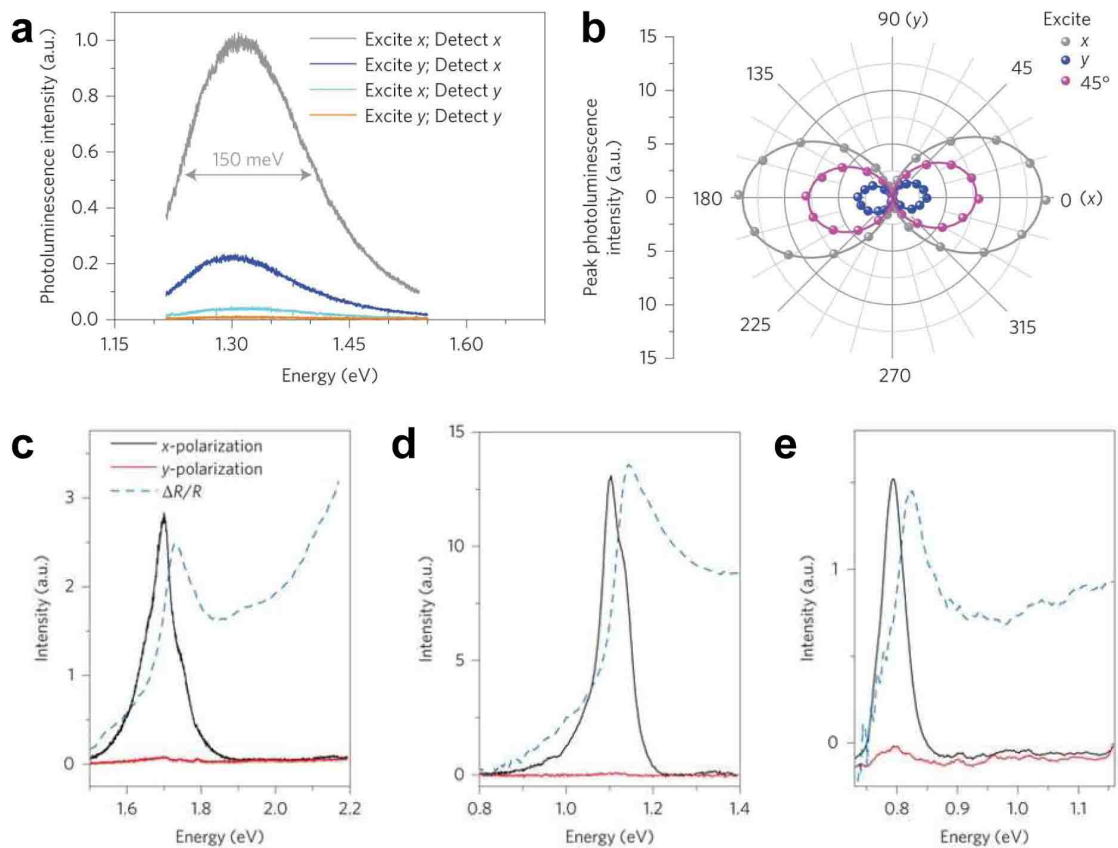

**Figure 3.** Polarized PL in BP. a) Polarization-resolved PL spectra of monolayer BP, with a laser excitation of 532 nm. The excitation and detection polarization are aligned in either the *x* (AC) or *y* (ZZ) directions, respectively. b) PL peak intensity as a function of the detection angle for an excitation laser polarized along the *x* (gray), 45° (magenta) and *y* (blue) directions. Solid lines are fitted curves using a $\cos^2\theta$ function ($\theta$ denotes the angle between the *x* axis and the detection angle). Reproduced with permission.[48] Copyright 2015, Springer Nature. PL spectra of c) monolayer, d) bilayer and e) trilayer BP at 77 K with unpolarized photoexcitation at 532 nm. Strong peaks are observed for PL detection under *x*-polarization (black curves). The peak energy matches well with that in the absorption spectra (blue dashed curves), confirming the direct nature of the bandgaps. PL is absent with *y*-polarization detection (red curves). Reproduced with permission.[32] Copyright 2017, Springer Nature.

## 2.3 Polarized Photodetection

Photodetectors can convert photons into free electrons, which are vital to modern photonics and optoelectronics. Generally, the responsivity (*R*), response time (*t*) and frequency response range are key parameters for describing the device performance.

Over more than a decade, a tremendous amount of studies on photodetectors based on 2D materials, such as graphene and TMDCs, have been published.[132] However, these detectors suffer from inherent drawbacks. For graphene, the zero-gap character leads to a high dark current. For TMDCs, the optical response is mainly limited to the visible frequency range. For BP, with a sizable and layer tunable direct bandgap (0.3-1.7 eV), the optical response can be naturally extended into the technologically important IR range, in which wide applications are involved, such as optical communications and thermal imaging. Most of the all, the intrinsic nature of linear dichroism in BP offers us an additional degree of freedom to detect and control the light polarization.

Experimentally, many kinds of photodetectors based on BP have been reported, working either in the photoconductive or photovoltaic mode.[133-145] Most of these focus on the visible range, with only a few focusing on the mid-IR range. Some typical studies will be briefly introduced here. An early work in 2014 reported a photodetector based on atomically thin BP films. The device showed a broadband (visible to near-IR) and relatively fast (~1 ms) response with a low responsivity of 4.8 mA/W.[139] Shortly after, the same group demonstrated light detection at a zero bias in locally defined BP PN junctions due to the photovoltaic effect.[136] Huang et al. reported a record-high responsivity in a back-gated BP photodetector in a high vacuum in the range of 400-900 nm.[141] They systematically explored the photoresponse dependence on laser power, temperature and source-drain channel length. A record-high responsivity was achieved, up to $7 \times 10^6$ A/W (20 K) and $4.3 \times 10^6$ A/W (300 K), with a channel length of 100 nm and laser power of 50 μW/cm$^2$, which was significantly higher than that in the previously reported BP-based devices. The key to achieving such a high responsivity is the low laser power and short channel length. They showed that the responsivity ($R$) significantly decreases with an increase in the laser power ($P$) and channel length ($L$), approximatively scaling as $R\sim1/P$ and $R\sim1/L^2$, respectively. The excellent device performance demonstrated by Huang et al. makes BP among the most promising candidates for future optoelectronic applications.

Most of previously reported BP-based photodetectors work in the visible to near-IR range. However, it is more attractive and practical to work in the mid-IR range. It is also worth noting that the capability to detect mid-IR light for bulk BP was demonstrated in the 1990s.[146, 147] After the reintroduction of BP as a layered IR material, Guo et al. reported the pioneering work of mid-IR photodetectors at 3.39 μm based on thin BP films (thickness: ~10 nm) in 2016, showing a relatively high responsivity of 82 A/W, even at room temperature.[137] Moreover, they demonstrated that the BP photodetector is sensitive to light polarization, which adds another degree of freedom to the device. More interestingly, the frequency response range can be further extended into the longer wavelength IR range in gated or strained devices. Chen et al. demonstrated that in a dual-gated BP photodetector (thickness: 5 nm), the photoresponse can be dynamically tuned from 3.7 to beyond 7.7 μm.[42] This can be well explained by the Stark effect, i.e., the vertical electric field across the BP layers shrinks the bandgap. The results unravel the potential of BP as a highly tunable and broadband mid-IR material. Additionally, an extended photoresponse beyond the fundamental bandgap is also expected in strained BP photodetectors, although there has yet to be an experimental demonstration. As reported previously, compressive strain can effectively reduce the bandgap of BP. A rough estimation shows that a 3% compressive strain may fully close the bandgap of bulk BP,[31, 35] leading to a semiconductor-to-mental transition, with the photoresponse expected to be further extended into the far-IR range.

Compared with graphene and TMDCs, the advantage of BP is its polarization sensitivity, which has been demonstrated by some groups in a broad spectral range.[134, 135, 137, 145] Two typical works related to the polarized photodetection in BP will be reviewed briefly here. Yuan et al. reported an early work to demonstrate the capability of polarized light detection in BP.[134] As illustrated in **Figure 4**a, they designed a ring-shaped metal electrode for photocurrent collection in order to avoid the influence from the shape and orientation of electrodes. Figure 4b shows the full spatial mapping of the photocurrent around the ring-shaped electrode at a wavelength of 1.5 μm. Clearly the device shows a strong polarization dependence, with the strongest photocurrent obtained for AC polarization and an almost zero photocurrent for ZZ polarization. Moreover, the contrast ratio of photocurrents along the two perpendicular polarizations can be further enhanced by electrical gating. Another interesting work by Bullock et al. demonstrates the first bias-selectable polarized photodetector with an ingenious design.[135] The device consists of two vertically stacked BP layers with their crystal orientations perpendicular to one another, with each acting as an isolated hole contact. The sandwiched MoS$_2$ layer serves as an electron contact (Figure 4c). As shown in Figure 4d, strong optical anisotropy was observed in the mid-IR range, and the extinction ratio $r_e$ (=$I_{AC}$/$I_{ZZ}$) reached as high as ~100 for both of the top and bottom BP layers under 3.5 μm light illumination, demonstrating a very large linear dichroism. Most interestingly, under a negative or positive bias, only 0° or 90° linearly polarized light can be detected, respectively, as shown in Figure 4e. This makes BP a promising candidate for future applications in polarized mid-IR detection.

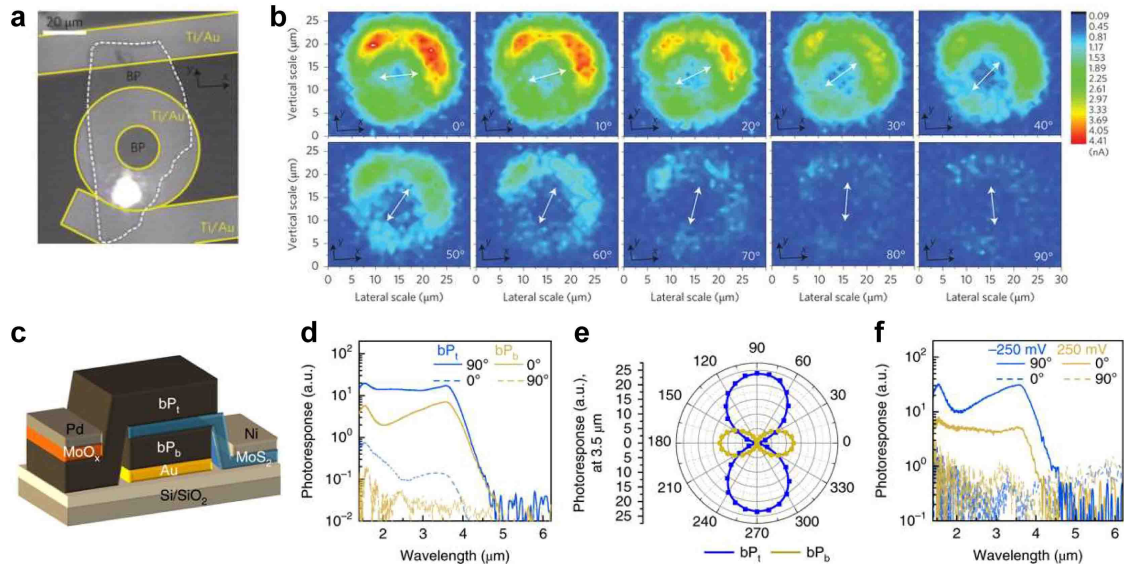

**Figure 4.** Polarized photodetection in BP. a) Optical image of a BP photodetector with a ring-shaped electrode for photocurrent collecting. b) Photocurrent mapping of the BP inside the inner ring under 1.5 μm light illumination with different polarizations. 0° and 90° denote the AC and ZZ directions, respectively. Reproduced with permission.[134] Copyright 2015, Springer Nature. c) Schematic illustration of BP/MoS$_2$ heterojunction photodiode, consisting of two vertically stacked BP layers with their crystal orientations perpendicular to one another. d) Spectrally resolved photoresponse measured from top BP (bP$_t$) and bottom BP (bP$_b$) with light polarizations normal and perpendicular to the device. e) Measured photoresponse under 3 μm light illumination in bP$_t$ and bP$_b$ as a function of the polarization angle. f) Spectrally resolved photoresponse of the device in the bias-selectable mode under four different conditions: ±250 mV for light polarization aligned to the AC direction of the top and bottom devices. Reproduced with permission.[135] Copyright 2018, Springer Nature.

**2.4 Anisotropic Ultrafast and Nonlinear Optical Properties**

In addition to the extensively studied steady-state and linear optical properties of BP, strong anisotropy has also been demonstrated in the transient and nonlinear optical properties. By means of scanning ultrafast electron microscopy (SUEM), Liao et al. achieved a direct visualization of the photoexcited carrier dynamics on the bulk BP surface in both space and time.[148] **Figure 5**a presents the SUEM images of carrier diffusion taken at different times after photoexcitation, clearly demonstrating the strong in-plane anisotropy: the carriers diffuse much faster in the AC direction than in the ZZ direction, regardless of the initial excitation polarization. Their results directly uncover the anisotropic carrier dynamics at nonequilibrium conditions in BP. They further showed that this anisotropy is much larger than that for carriers at near-equilibrium conditions.

Ge et al. used optical pump-optical probe spectroscopy to probe the anisotropic carrier dynamics in bulk BP.[149] Strong anisotropy in transient absorption was observed at a time scale of ~1000 ps. Regardless of the probe polarization, the transient absorption is strongly dependent on the pump polarization, similar to that for the case of steady-state absorption. Later, the same group further explored the effect of magnetic fields of up to 9T on the anisotropic transient optical response in bulk BP.[150] They found that the degree of anisotropy was significantly degraded by the magnetic fields. This can be explained by the fact that the photoexcited carriers undergo cyclotron motions in magnetic fields. During this process, the wavefunctions with momenta along different directions begin to mix, and hence the optical response of BP tends to become less anisotropic. He et al. also demonstrated strong anisotropic transport properties for photoexcited carriers in bulk BP using optical pump-optical probe spectroscopy, and showed that the carrier mobilities (for both holes and electrons) in the AC direction are much larger than that in the ZZ direction.[151]

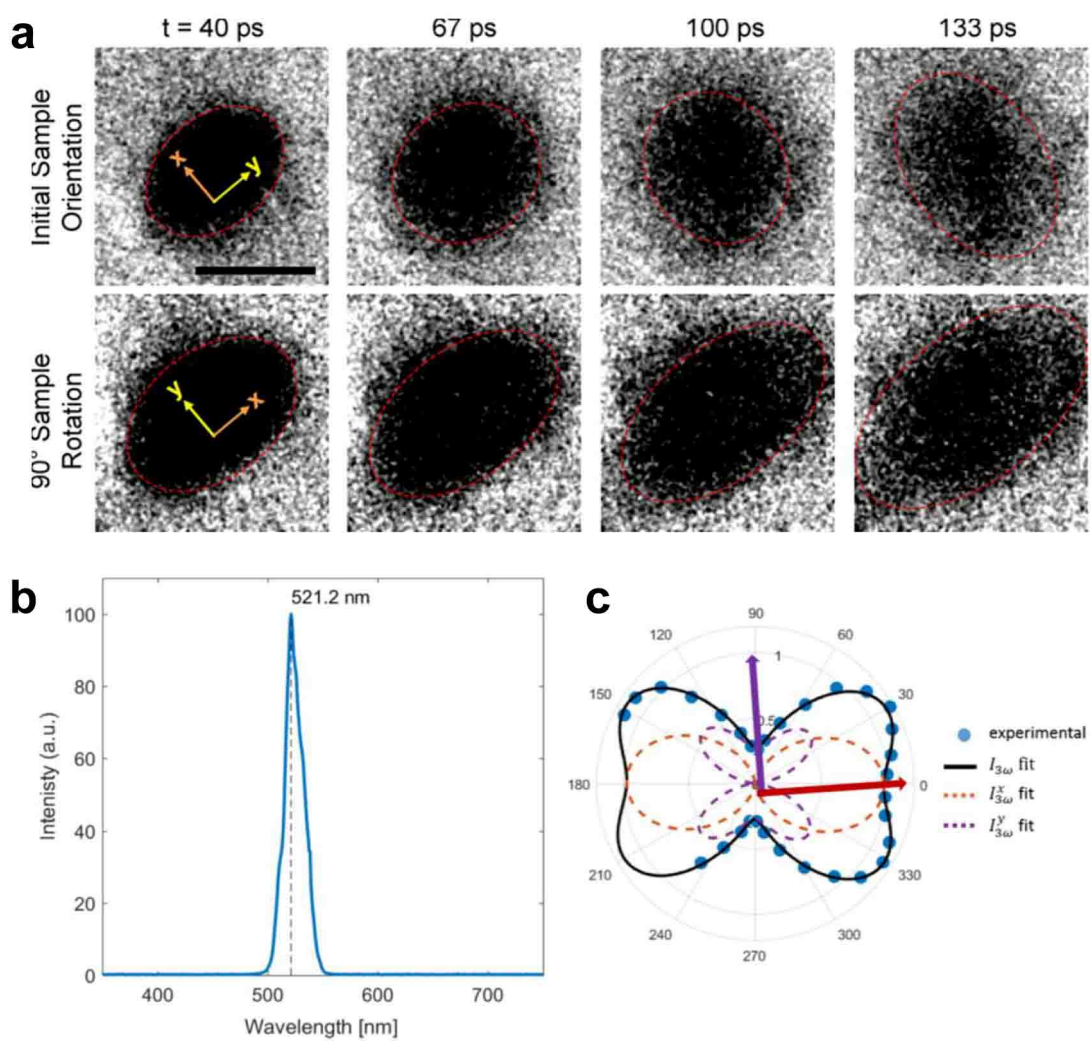

**Figure 5.** Anisotropic transient and nonlinear optical properties of BP. a) SUEM imaging of hole diffusion in bulk BP after photoexcitation, clearly showing that the holes preferentially diffuse along the *x* (AC) direction. Scale bar: 60 μm. Reproduced with permission.[148] Copyright 2017, American Chemical Society. b) THG spectrum of a BP flake (thickness: 30 nm), with laser excitation at a wavelength of ∼1560 nm. c) Polarization dependence of the THG intensity of a BP flake (thickness: 26 nm). The red and purple arrows denote the AC and ZZ directions, respectively. The curves are fit to the data using Equation 2. Reproduced with permission.[152] Copyright 2017, American Chemical Society.

Nonlinear optical properties have been widely investigated in 2D materials, such as TMDCs and hBN. Kumbhakar et al. studied the multiphoton absorption properties of hBN by the *z*-scan technique,[153] and later reported a significant enhancement in hBN-graphene oxide heterostructure compared with the bare hBN.[154] Second-harmonic generation (SHG) is a typical second-order nonlinear optical process, which has been widely observed in 2D materials, such as noncentrosymmetric TMDCs and hBN,[155-157] while for BP, SHG is absent due to its centrosymmetric crystal structure. Only third-order nonlinearity is permitted, such as third-harmonic generation (THG) and four wave mixing (FWM). Experimentally, THG has already been observed in BP by a few groups.[152, 158, 159] A typical example is shown in Figure 5b. As expected, it was determined that the THG intensity is highly anisotropic as well, strongly depending on the laser polarization. In detail, the THG intensity ($I_{3\omega}$) can be expressed as follows:

$$I_{3\omega} \propto I_\omega^3 \left[ \left( \chi_{xx}^{(3)} \cos^3\theta + 3\chi_{xy}^{(3)} \cos\theta \sin^2\theta \right)^2 + \left( \chi_{yy}^{(3)} \sin^3\theta + 3\chi_{yx}^{(3)} \sin\theta \cos^2\theta \right)^2 \right] \quad (2)$$

where $I_\omega$ is the intensity of the excitation laser, $\theta$ is the polarization angle, and $\chi$ denotes the third-order susceptibility tensor, with the subscripts *x* and *y* referring to the AC and ZZ directions, respectively. As shown in Figure 5c, the THG intensity can be well fitted by Equation 2. This clearly shows that the contribution from the AC-component exhibits a two-fold pattern (red dashed curve), and the ZZ-component contributes a four-fold pattern (purple dashed curve). The overall result is that the total THG (black solid curve) exhibits a four-fold polarization dependence, but the maximal intensity does not occur in the AC direction. In addition, Rodrigues et al. reported the preliminary observation of resonantly enhanced THG in a thin BP flake, in which the excitonic resonances nearly match the energy of the excitation laser.[158] This may stimulate further experimental research on the nonlinear optical response of few-layer BP, which hosts a QW-like subband structure as mentioned above. When the excitation energy is in resonance with one pair of subbands, the optical nonlinearity is expected to be significantly enhanced.

## 3. Plasmons in Anisotropic 2D Materials

The discovery of electrical tunable and highly confined plasmons in graphene have stimulated considerable efforts to explore new plasmonic 2D materials.[27, 104, 105, 160] Plasmons in 2D materials with in-plane anisotropy have recently attracted increasing attention due to the possibility of realizing in-plane hyperbolic plasmon polaritons, the iso-frequency contour of which is a hyperbola.[104] Such special dispersion topology supports directional propagating polaritons with divergent densities of states, which enables a series of potential applications in nanophotonics.[96-99] However, no experiment is reported to have observed the hyperbolic plasmons in the natural 2D vdWs surface. The in-plane hyperbolic phonon polaritons, on the other hand, have been demonstrated in the natural surface of $MoO_3$[161, 162] and the structured h-BN[163] surfaces. Recently, BP, as a typical in-plane anisotropic 2D material, has been proposed to sustain the natural in-plane hyperbolic plasmons. The large anisotropy existing in the intraband and interband transitions along the two principle axes is considered to be the key to the hyperbolic regime. The high carrier mobility of approximately 1000 $cm^2/(V·s)$ is suitable for the observation of plasmons with longer lifetimes. In addition, the hyperbolic regime in BP films can be tuned by modifying the carrier densities and band gap by gating, doping and straining, which is not viable in the metamaterials based on noble metals. Therefore, BP films are treated as a model system to host the anisotropic and hyperbolic 2D plasmons.[104] Great theoretical[74-95] and experimental[106-109] efforts have been devoted to investigating the plasmons in this natural anisotropic material. Below we discuss the theoretical studies on the plasmon dynamics of BP films and the experimental progress in detecting BP plasmons, as well as the potential applications of plasmons in anisotropic 2D materials. At the end, we give a perspective on other anisotropic 2D materials for plasmons (focusing on anisotropic layered semimetals).

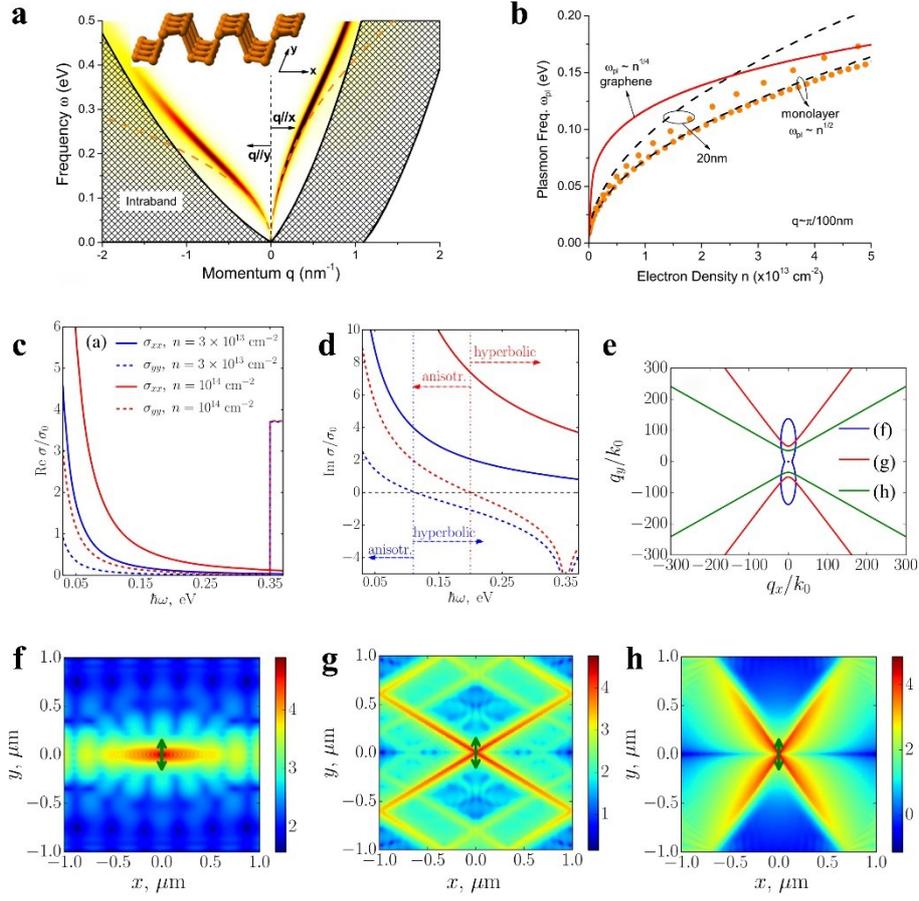

**Figure 6.** Anisotropic and hyperbolic plasmons in mono- and multilayer BP. a) Plasmon dispersion calculated by the loss function of monolayer BP at 300 K with electron doping of 1 × $10^{13}$ cm$^{-2}$, with momentum $q$ along the two principle axes $x$ (AC) and $y$ (ZZ). The shaded areas indicate the Landau damping regions. b) Plasmon dispersion for the monolayer and a 20 nm BP thick film as a function of the carrier concentration with $q$ along $x$. Graphene plasmons are plotted for comparison. Dashed black lines show the $n^{1/2}$ scaling. Reproduced with permission.[74] Copyright 2014, American Physical Society. Calculated real c) and imaginary d) parts of the conductivity of BP thin films along the two principle axes $x$ (AC) and $y$ (ZZ), for different carrier densities. e) Iso-frequency contours of plasmons with conductivity given by c) and d), with $n = 10^{14}$ cm$^{-2}$, $E = 0.165$ eV for the blue line, $n = 10^{14}$ cm$^{-2}$, $E = 0.3$ eV for the red line and $n = 10^{14}$ cm$^{-2}$, $E = 0.165$ eV for the green line. f) to h) The electric field distribution in real space (log scale) for surface plasmons excited by a $y$-polarized electric dipole. The corresponding $q$-space dispersions are shown in e). Reproduced with permission.[79] Copyright 2016, American Physical Society.

## 3.1. Hyperbolic Plasmons in Anisotropic 2D Materials

The topology of 2D plasmons is determined by the sheet conductivity tensor with the form $\begin{bmatrix} \sigma_{xx} & \sigma_{xy} \\ \sigma_{yx} & \sigma_{yy} \end{bmatrix}$.[104] Here, $x$ and $y$ are the directions along the principle axes, and $\sigma_{xy} = \sigma_{yx} = 0$ in nonmagnetic systems. In graphene, the conductivity is isotropic with $\sigma_{xx} = \sigma_{yy}$, leading to an isotropic dispersion of plasmon, while in anisotropic 2D materials, such as BP, the conductivity along the two principle axes have different values. Therefore, anisotropic or even hyperbolic ($\sigma''_{xx}\sigma''_{yy} < 0$) plasmon polaritons are sustained.[79, 104] Such topology of 2D plasmons can be described by the simplified plasmon dispersion relation in the 2D case[79] by assuming that the damping rate is low ($\sigma = i\sigma''$) and that $q_x, q_y \gg \omega/c$ as follows:

$$\frac{q_x^2}{\sigma''_{yy}} + \frac{q_y^2}{\sigma''_{xx}} = 2|q|\omega\left(\frac{\varepsilon_0}{\sigma''_{xx}\sigma''_{yy}} - \frac{\mu_0}{4}\right) \tag{3}$$

where $|q| = \sqrt{q_x^2 + q_y^2}$, and $\varepsilon_0$, $\mu_0$ are the permittivity and permeability in a vacuum, respectively.

In the purely anisotropic case ($\sigma''_{xx} > 0$, $\sigma''_{yy} > 0$), the iso-frequency contour of the plasmon is an ellipse, whose long axis is along the crystal direction with a smaller conductivity (the blue line in **Figure 6**e). The plasmon energy, on the contrary, transmits preferentially along the other axis, as shown in Figure 6f. In the hyperbolic case ($\sigma''_{xx}\sigma''_{yy} < 0$), the plasmon dispersion in $q$ space is a hyperbola (green and red lines in Figure 6e). This gives rise to directional propagating polaritons in Figure 6g and 6h, with the direction of the plasmon beams determined by the following:[79, 164]

$$y = \pm\sqrt{|\sigma''_{xx}/\sigma''_{yy}|}\, x \tag{4}$$

Thus, the propagating directions of plasmons in the hyperbolic regime are nearly identical for all momenta, which make the plasmons more localized than those in the elliptic regime.[79, 99]

Due to the highly anisotropic in-plane dynamics, a number of theoretical works were performed to study the plasmons in BP.[74, 75, 77, 79, 88] Low et al. studied collective electron excitations in monolayer and multilayer BP within the random phase approximation (RPA) using an effective low-energy Hamiltonian.[74] As mentioned above, the optical properties in BP evolve with the layer number, leading to a thickness-dependent optical conductivity.[31, 32] In monolayer BP, due to the anisotropic effective mass, the plasmon dispersions are also anisotropic along the two principle axes, as indicated by the loss function results (color plots) in Figure 6a. The loss function quantifies the spectral weight of the plasmons. Because of the large bandgap (above 1 eV) in the monolayer BP, the low energy plasmon dispersion is mainly determined by the intraband transitions and follows the $\sqrt{q}$ relation (dashed lines), as in conventional 2DEGs, while the plasmon dispersion at higher energy deviates from the free carrier case as a result of the intraband Landau damping. The plasmons in multilayer BP are more interesting because of the increased mobility and, more importantly, the increased intraband and interband coupling. As shown in Figure 6b, the scaling relation of the plasmon frequency ($\omega$) in monolayer BP and the carrier density ($n$) is $\omega \propto n^{1/2}$, as in 2DEGs, while for thicker films, they found that $\omega \propto n^{\beta}$ with $\beta < \frac{1}{2}$. The nonparabolicity reflects the modification of plasmon dispersion by the strong interband coupling in multilayer BP. Margulis et al. used the tight-binding model to study the surface plasmon polaritons in a monolayer BP between two different dielectric surroundings.[88] In the doped BP, three types of localized plasmon modes are hosted:

the transverse magnetic mode, the transverse electric mode and the hybridization mode between them (TE, TM and TEM modes). The TE and TM modes (TEM mode) only exist for the wave vector along (away from) the principle axes.

In 2016, Nemilentsau et al. examined the optical conductivity in multilayer BP and theoretically predicted the existence of hyperbolic plasmons.[79] By considering both the intraband and interband contributions, the conductivity can be given by the following:

$$\sigma_{kk} = \frac{ie^2}{\omega + i\eta} \frac{n}{m_k} + s_k \left[ \Theta(\omega - \omega_k) + \frac{i}{\omega} \ln \left| \frac{\omega - \omega_k}{\omega + \omega_k} \right| \right] \quad (5)$$

where $k$ denotes the crystal axes $x$ (AC) and $y$ (ZZ), $n$ is the sheet carrier density of the electrons, $\eta$ is the relaxation time, $m_k$ is the effective mass of the electron, $\omega_k$ is the frequency of the bandgap along the $k$ direction, and $s_k$ represents the intensity of the interband transitions. The first term of Equation 5 describes the Drude component of conductivity, which is responsible for the positive value of $\sigma_{kk}''$ at the low energy side in Figure 6d, while the second term, the interband transition component, gives rise to the negative value at the higher energy side in Figure 6d. Due to the crystal asymmetry, the sign change of $\sigma_{kk}''$ can, in principle, occur at different energies, leading to an energy interval for hyperbolic plasmons where $\sigma_{xx}''\sigma_{yy}'' < 0$. Such hyperbolicity naturally exists in the anisotropic 2D materials, which do not need artificial structures as in metasurfaces. In addition, the carrier density and band gap of BP can be tuned in situ by the strain and electrical method. For example, due to the stark effect, Kim et al. successfully tuned the band gap of BP in a wide range, from a semiconductor with a moderate bandgap to an anisotropic Dirac semimetal via the in situ surface doping of potassium (K) atom.[41] Dynamic tuning of the carrier density and bandgap has also been reported in thin BP films by the electrical gating method.[43] Since the sigma-near-zero point ($\sigma'' = 0$) is determined by the coupling between the positive part from intraband transitions and the negative part from interband transitions, the topological transition frequency can be tuned by modifying the carrier density and band gap. For example, by increasing the carrier density, the strength of the intraband transition is increased, leading to an enhancement of the positive value of the imaginary part of the conductivity. Given that the interband transitions remain the same, the sigma-near-zero points along both axes therefore move to higher energies, along with the simultaneously change in the $\sigma''$ ratio, as shown in Figure 6c and 6d. According to Equation 4, the travelling direction of the hyperbolic plasmon is dependent on the $\sigma''$ ratio along the two axes, enabling in situ manipulation of the propagation of the hyperbolic plasmon beams (discussed below).

The above analysis of hyperbolic plasmons are based on the local assumption ($q \rightarrow 0$), where $\sigma$ is a function of the only variable $\omega$. However, in the thin film of BP, the wavelength of plasmon polaritons in the hyperbolic regime is extremely confined ($q \gg k_0$) (Figure 6e), leading to the case where the nonlocal effect should be considered.[77, 83, 94, 95] Correas-Serrano et al. analyzed the anisotropic optical conductivity of BP and took a nonzero wave vector into account.[77] The topology of plasmons can be predicted more accurately by considering the wave vector-dependent conductivity. **Figure 7**a to 7f show the nonlocal conductivity results for a 10 nm BP film at a frequency corresponding to the hyperbolic regime in the local assumption. For small $q$, the conductivities are uniform and converge to the local values. At large $q$, the conductivities show $q$-dependent dispersions, the dispersion rates of which are anisotropic along the two principle axes. In addition, the nonlocal effect at a large $q$ may even cause the conductivity components to change from metallic to dielectric or vice versa. Figure 7g to 7j show the comparison of iso-frequency contours of plasmon dispersion with local and nonlocal assumptions. For the elliptical plasmons (panels (g) and (h)), where the wavenumbers are relatively low, nonlocality has a moderate effect on the plasmon dispersion, while the effect of nonlocality is much more pronounced in the iso-frequency contours of the hyperbolic case (panels (i) and (j)), in which the hyperbolic dispersion is replaced by an almost perfect canalization along $y$ due to the unusual nonlocality-induced dielectric-metallic transition. The effect of nonlocality can also be reflected by the modification of the spontaneous enhanced resonance (SER) of sources near the surface plasmons. Figure 7k and 7l present the SER intensity of a $z$-polarized dipole as a function of its distance to the BP films with local and nonlocal models. In panel (l), where the plasmon dispersion is elliptic in the local assumption, the local model provides good accuracy due to the small effect that nonlocality has on the plasmon dispersion, while in the hyperbolic regime (panel (m)), the disagreement between the nonlocal and local models is more pronounced due to the large $q$ value. In contrast, the SERs of graphene in the local and nonlocal models show similar trends due to relatively small *q* values. Moreover, an investigation was performed by Petersen et al. to explore the nonlocal effect on the plasmon dispersion and its corresponding Purcell factor in activated BP thin films by either electron doping or optical pumping.[83] They demonstrated that the Purcell factor can be changed by orders of magnitude by considering the nonlocal effects when the distance between the dipole and the surface is small (below 5 nm).

In addition to the analysis of the intrinsic plasmons in BP films, many works have been conducted to simulate the plasmon modes in the case of practical experiments.[75, 78, 80, 85, 87, 91-93] For example, magnetoplasmons of BP multilayers under a magnetic field, localized plasmon resonance modes and edge plasmon modes in the nanostructures of BP films and the effect of the surrounding dielectric environment are investigated in theory, where intrinsically and highly anisotropic plasmon modes are discovered due to the anisotropic effective mass along the two principle axes.[78, 85, 91-93] Lee et al. studied the plasmon dynamics of BP film by placing it on top of a conducting periodic array, in which anisotropic acoustic plasmons were revealed along the AC and ZZ directions.[87] Because of the different plasmon confinement, the plasmon dispersions along the two axes were found to follow linear scaling in the different spectral ranges. Ghosh et al. explored the anisotropic electron energy-loss spectra in monolayer BP, based on ab initio time-dependent density-functional-theory calculations.[80] At low energies or long wavelengths, intraband plasmon appears with doping with a frequency of $\sqrt{q}$ scaling. At a high energy above the band gap, plasmons are dominated by the interband transitions.

Since BP is air sensitive, the effect of the degradation-induced defects and impurities on the plasmons needs to be considered. The disorders increase the scattering rate, which reduces the lifetime of the plasmons. Jin et al. studied the disorder effect on the lifetime of the plasmon excitations for single-layer and bilayer BP. It was determined that the long range disorder damps the plasmon modes in a stronger way than the local point defects due to the long wavelength nature of the collective excitations and that the disorders have a stronger effect on the out-of-phase plasmon mode in bilayer BP.[75]

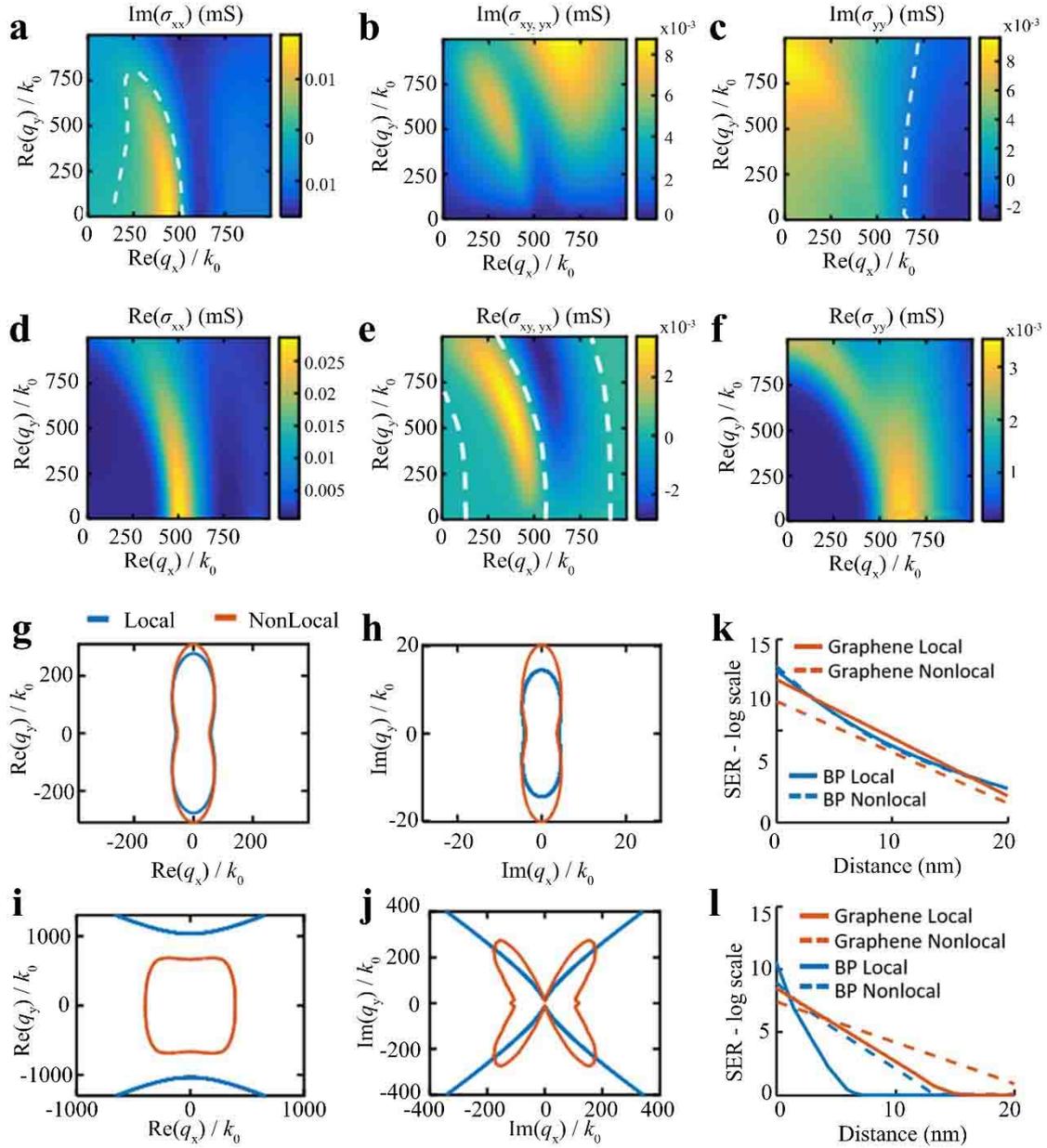

**Figure 7.** Nonlocal effect on the plasmon dispersion of BP. a) to f) Nonlocal conductivity components of BP versus real in-plane wavenumbers along the two crystal axes $x$ (AC) and $y$ (ZZ), operating at a frequency where the plasmon is in the hyperbolic regime by local assumption. g) to j) Comparison between the iso-frequency contours of plasmons with and without considering the nonlocal effect. k) and l) Spontaneous emission rate of a $z$-oriented point source as a function of its distance to BP films and graphene sheets, considering local and nonlocal effects. Reproduced with permission.[77] Copyright 2016, IOP Publishing.

**3.2. Experimental Progress for Plasmons in Anisotropic 2D Materials**

In contrast to the large number of theoretical studies of plasmons in anisotropic 2D materials, the experimental reports are rather limited. One of the reasons for this is the low carrier concentration in BP, leading to the plasmon wave vector being much higher than that in graphene.[77, 79] This adds difficulties in detecting the anisotropic plasmons by both far-field and near-field methods, since the resonator size and the plasmon wavelength are inversely proportional to the plasmon wave vectors.[105] The second reason is that BP layers are quite unstable, with the sample quality severely degraded under ambient conditions. The impurities and defects induced by the degradation increase the damping channels for plasmons, which further challenges people in experimental observation and tuning of plasmons in this material.

To compensate the former, one method is to increase the carrier density via chemical doping. Huang et al.[107] reported the study of plasmons in nanostructured black phosphorus carbide (b-PC) films (**Figure 8**a), synthesized via the carbon doping technique. As shown in Figure 8b, two peaks can be observed in the extinction spectra, the frequencies of which are dependent on the ribbon width. The two peaks are attributed to the hybrid modes from the plasmon-phonon coupling. In addition, such modes can be tuned via electrical gating (Figure 8c) and changing the ribbon orientation (Figure 8d). The latter reflects the intrinsic in-plane anisotropy of b-PC, offering a new degree of freedom for manipulating the plasmons, in contrast to the isotropic nature of graphene.

Due to the semiconducting nature of BP, the plasmon resonance frequency and amplitude can also be enhanced by ultrafast laser pulse pumping. Huber et al.[108] studied the plasmons in the $SiO_2$/BP/$SiO_2$ heterostructure using scanning near-field optical microscopy (s-SNOM) combined with the ultrafast pump/probe scattering-technique, as shown in Figure 8e. Hybrid plasmon-phonon excitations are observed at approximately 34 THz after the illumination of near-infrared pump pulses, which exhibit a well-defined frequency, momentum and an excellent coherence. As shown in Figure 8f, clear interference fringes emerge even within the rising time of the scattering response. Such fringes are standing wave interference patterns between the stimulating polariton wave and the wave from the edge reflection, which are hallmarks of the excitation of propagating polaritons. The average amplitude of these fringes rises from zero to above the noise level on the order of < 50 fs. Fringes can be clearly observed at all times during the lifetime of the excited polaritons and disappear after approximately 5 ps, which demonstrates an ultrafast switching. The polariton activation process corresponds to the excitation of electron-hole pairs with intraband relaxation in the subpicosecond time scale, where the subsequent process is determined by the recombination of the interband electron-hole pairs with a longer relaxation time. These transient hybrid modes have the potential for applications in ultrafast nanophotonic devices due to their high switching contrast, fast switching speed, excellent coherence character and constant wavelength.

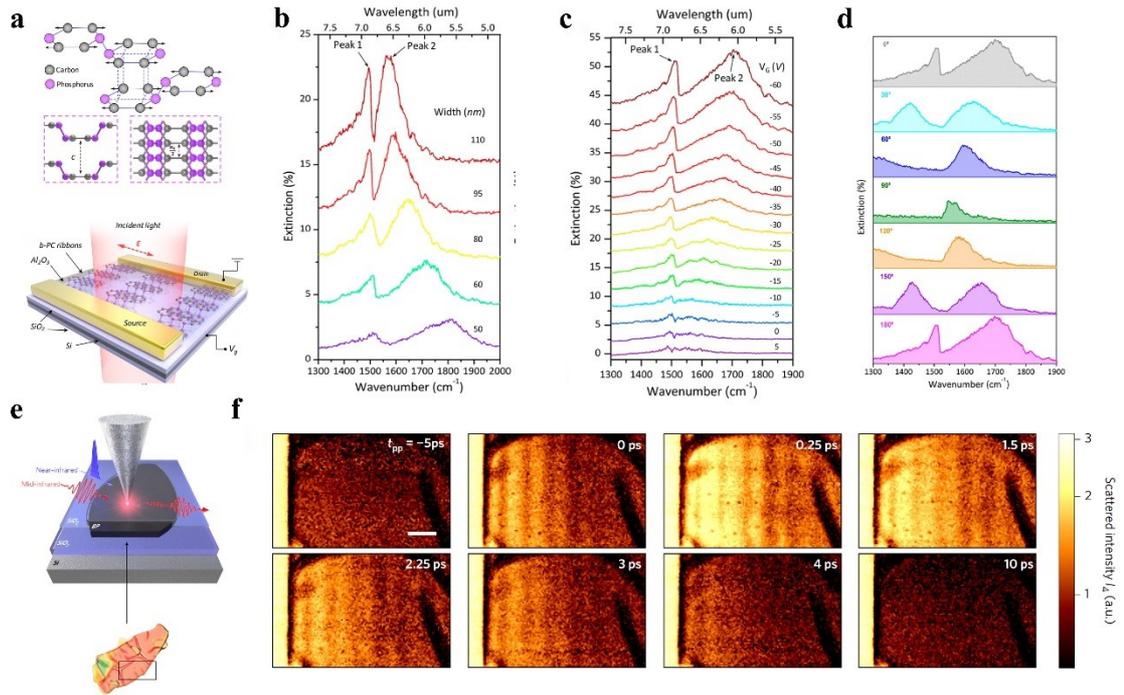

**Figure 8.** Experimental observation of plasmon modes in doped BP films and $SiO_2\backslash BP\backslash SiO_2$ heterostructures. a) Schematics of the crystal structure of black phosphorus carbide (b-PC) and the setup for mid-IR transmission measurements in nanoribbon arrays via FTIR microscopy. b) to c) Width and gate dependence of the extinction spectra with the ribbon along the ZZ direction. d) Polarization-dependent extinction spectra. Reproduced with permission.[107] Copyright 2018, American Chemical Society. e) Upper panel: Schematic of the setup for measuring hybrid plasmon-phonon polaritons by near-field microscopy combined with the ultrafast near-infrared pump/probe technique. Lower panel, an optical image of the measured BP flake between $SiO_2$ on the top and bottom. f) The scattered near-field intensity of the heterostructure measured with different delay times between the pump and probe pulses. Reproduced with permission.[108] Copyright 2016, Springer Nature.

In addition to the above-mentioned experiments, other efforts have also been devoted to observing the plasmon modes in BP. Abate et al. reported the observation of an outside edge fringe in BP thin films by the near field microscopy, which was attributed to the formation of a new conducting layer at the BP surface.[109] The anisotropic interband plasmons at ultraviolet energies were observed by Nicotra et al.

through momentum-resolved EELS coupled with scanning transmission electron microscopy.[106]

### 3.3. Potential Applications of Plasmons in Anisotropic 2D Materials

Because of their unique plasmon dispersion, the hyperbolic plasmons can propagate in a certain direction with a large density of states and higher filed confinement, leading to applications of spontaneous radiation enhancement,[99, 164, 165] hyperlens,[166-169] negative index materials[99, 170-173] and thermal management.[174-181] Many of these have been realized in hyperbolic metasurfaces, created by artificial subwavelength structuring from visible to microwave frequency ranges.[170, 182, 183] However, in such artificially created hyperbolic surfaces, the wave vector of plasmon polaritons is limited by the inverse of the structure size, which hinders the potential of hyperbolic plasmons.[163] The anisotropic 2D materials present a natural platform for realizing the planar hyperbolic plasmons with much larger momentum,[161, 162] which may fulfill the applications with enhanced light-matter interaction[77, 83] and near-field radiative heat transfer[178, 180, 181] as a result of higher electromagnetic confinement and more diverging photonic density of states. More importantly, the in-plane electronic properties of 2D materials can be tuned both electrically and mechanically,[27, 100] adding a new degree of freedom to tune the 2D plasmons.[104, 105] Since there are already detailed reviews about the application of hyperbolic metasurfaces,[96, 98, 99] here we mainly focus on the potential applications of planar plasmons, which are unique to anisotropic 2D materials.

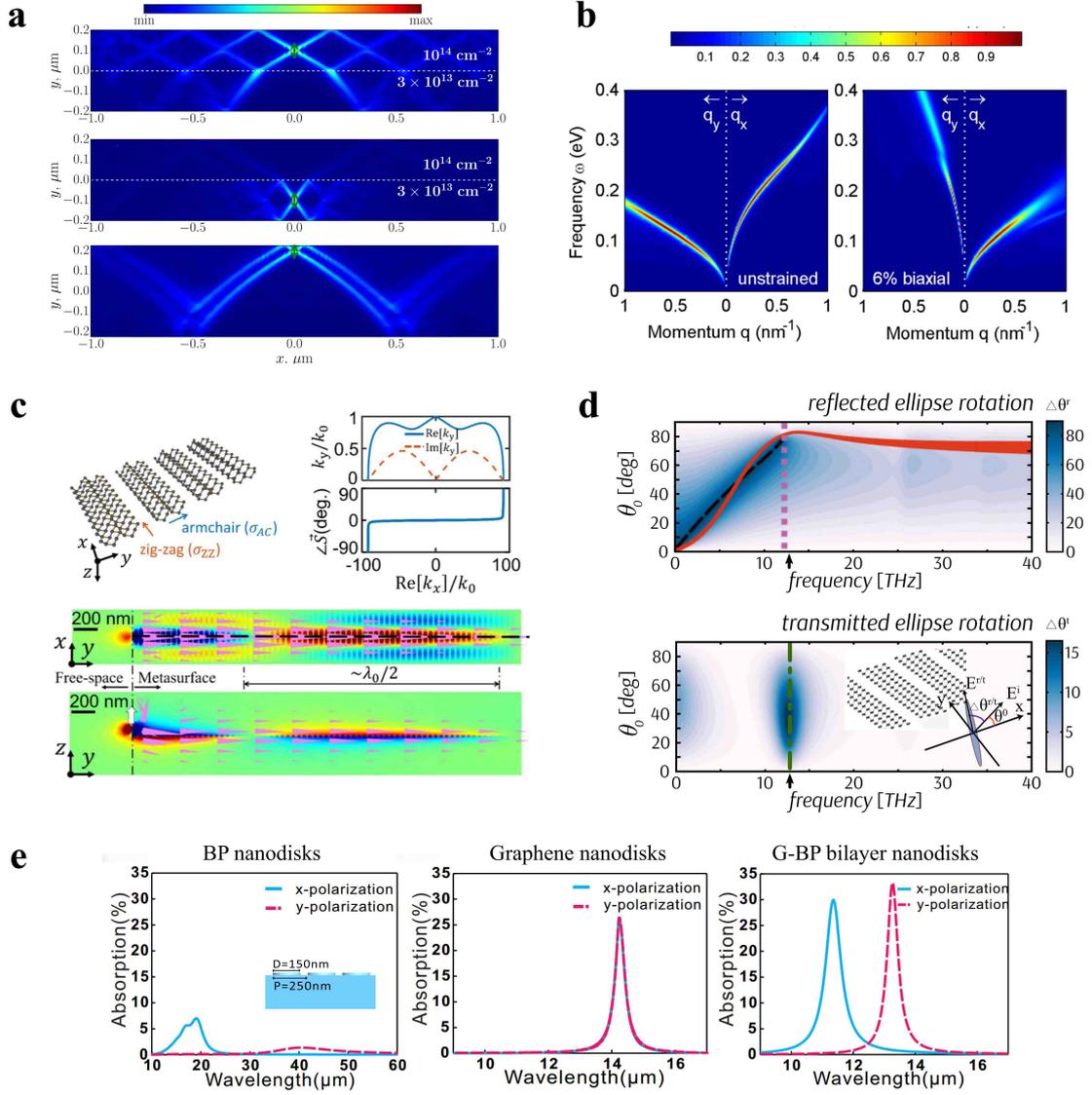

**Figure 9.** Potential applications for anisotropic plasmons in BP. a) Demonstration of the reflection, refraction and bending of plasmon propagation by modifying the carrier concentration in 2D materials. The parameters are the same as in Figure 6c and 6d. The plasmons are launched by a *y*-polarized electrical dipole (green arrows). The electron concentration in the lowest panel changes gradually from $n = 10^{14}$ cm$^{-2}$ (upper layer) to $n = 2 \times 10^{13}$ cm$^{-2}$ (lower layer). Reproduced with permission.[79] Copyright 2016, American Physical Society. b) Plasmon dispersion of unstrained and 6% biaxially strained monolayer BP along the *x* (AC) and *y* (ZZ) directions, calculated by the loss function. Reproduced with permission.[76] Copyright 2015, AIP Publishing. c) Plasmon collimation induced by loss in an array of BP nanoribbons with the ribbon direction along ZZ. The iso-frequency contour and the direction of the Poynting vector with respect to the *y* axis are calculated at the resonant frequency of the ribbon array with an effective conductivity of $\sigma_{yy}$ = 143 mS and $\sigma_{xx}$ = 0.5 + 57*i* μS. The two-color maps present the *x* component of the magnetic field, on which the Poynting vectors are plotted as arrows, excited by a *z*-polarized dipole (plotted as a white arrow in the bottom panel). Reproduced with permission.[94] Copyright 2017, American Physical Society. d) Rotation of the major-axis for the ellipse in the scattered fields of a microribbon array fabricated from anisotropic films. The ribbon direction is along the high-inductance axis (corresponding to the ZZ direction in BP). The plasmon resonance frequency is denoted by the black arrows. Reproduced with permission.[81] Copyright 2017, American Physical Society. e) Comparison of the plasmon resonance modes between a monolayer BP, a graphene sheet and a G-BP bilayer heterostructure. *x* and *y* are along the AC and ZZ directions of BP, respectively. Reproduced with permission.[86] Copyright 2018, Optical Society of America.

As discussed above, the propagating direction of hyperbolic plasmons is determined by the ratio of $\sigma''$ along the two principle axes, which can be tuned by gating. This paves a way for dynamic manipulation of the hyperbolic surface waves,[79, 94, 164] which is highly challenging in the metasurfaces based on noble metals. **Figure 9**a presents examples for the propagation control of hyperbolic plasmon rays.[79] The upper two panels are composed of two sheets placed next to one another with different electron concentrations. When plasmon waves are generated by the dipole located at the regime with a higher electron concentration, a majority of the plasmon's energy propagates through the interface to the lower concentration side. In contrast, the interface reflection of the plasmon dominates when the dipole is on the lower doped side. Moreover, we can achieve plasmon ray bending in a plane with a gradually varying doping across the *y* direction, as shown in the lower panel of Figure 9a.

The puckered structure of BP allows it to support a high level of strain, which can be used to modulate the band structure.[31, 35, 36] Thus, one can utilize strain to modify the plasmons in BP thin films.[76, 90] Lam et al. used the first principle method to study the plasmon dispersion of the monolayer BP under strain.[76] Figure 9b shows the loss functions for the plasmon dispersion of the monolayer BP with and without strain. By modifying the effective mass ratio, the application of strain can engineer the anisotropy of the plasmon dispersion along the two principle axes, leading to a higher dispersion rate along *x* (*y*) in the unstrained (6% biaxially strained) monolayer BP. Saberi-Pouya et al. studied the plasmon dynamics of monolayer, bilayer, and spatially separated double-layer BP under uniaxial train by calculating the effective masses along the two axes.[90] They found that the strain orientation is important for the plasmon response under strain. At a fixed strain direction, the plasmon dispersions exhibit different strain dependences along the two principle axes. These changes in plasmon dispersion with respect to the applied strain demonstrate the potential for the piezo-optic applications.

By exploiting the natural conductivity anisotropy, anisotropic 2D materials can add additional design flexibility in engineering metasurfaces. Take the plasmon canalization devices as an example.[77, 94, 99, 164] Previous implementations of such devices have always been based on modulation or arranging strips of isotropic materials, such as graphene[164] and silver,[170] to engineer metamaterials with suboptimal *σ*-near-zero topologies, in which the loss and the transverse conductivity component are important hindrances. The anisotropic 2D material provides an ideal platform to fulfill these conflicting conditions. Correas-Serrano et al.[94] proposed a metasurface based on BP films, which is composed of a subwavelength ribbon array to achieve loss-based canalization and collimation of plasmons, as shown in Figure 9c. A plasmon resonance mode of the metasurface is designed to give a high loss in the mid-IR, leading to an anisotropic ratio of the effective conductivity of over 2500. Such a high conductivity anisotropy exhibits exceptional capabilities for canalization as a result of a nearly flat iso-frequency contour and moderate loss (Figure 9c). This can be manifested by the magnetic field distributions (color maps) and the Poynting vector (arrows) in Figure 9c, which demonstrate a narrow and highly directional propagating plasmon beam confined to the surface.

It is known that the linear birefringent effect exists in anisotropic media, where the phase accumulation of light is polarization-dependent.[184] This effect in planar photonics can be realized in anisotropic metasurfaces through artificial manipulation of isotropic surfaces[185-189] or the unstructured natural films of anisotropic 2D materials.[81, 189, 190] The latter possesses elliptic or even hyperbolic plasmon regimes that can be treated as ideal platforms for ultrathin linearly birefringent retarders.[189] Because the linear birefringence and the polarization rotation of light are directly determined by the ratio of the scattered amplitudes, microribbon arrays fabricated from the homogeneous anisotropic 2D surface, where the scattered amplitude in the perpendicular direction of the ribbons can obtain significant enhancement by exiting the localized plasmons, were proposed in order to implement the plasmon-enhanced linear birefringence.[81] When light is scattered by such ultrathin metasurfaces, as shown in Figure 9d, the incident linear polarization can be converted into its orthogonal direction or even circular polarization. In addition, the amplitude ratio can also be modified by tuning the carrier density and effective masses through the electrical gating or strain, rendering the anisotropic ribbon array as a viable platform for tunable polarization converters.

The weak vdWs interlayer force in 2D materials makes it possible to isolate 2D films and restack them into arbitrary stacking heterojunctions without the restriction of the lattice match.[101-103] By stacking 2D materials with different functions, one can achieve new features that are not available in each individual component. Hong et al.[86] proposed a plasmonic heterostructure composed of a graphene-black phosphorus (G-BP) bilayer nanodisk array, as shown in Figure 9e. In the monolayer BP nanodisk array, the plasmon modes along the two principle axes have anisotropic frequencies due to the anisotropic effective mass. However, the amplitudes of the resonance modes are much lower than the plasmon modes in the graphene nanodisk array, while no anisotropy can be found in the latter. By combining the two layers into a heterostructure, the G-BP bilayer nanodisk array hosts anisotropic plasmon modes with strong plasmon absorptions along both axes at the same time. Such anisotropy is affected by the thickness of the insulating layer between the G-BP bilayer. Nong et al.[89] theoretically investigated the plasmon hybridization in the graphene-BP nanoribbon heterostructure. The coupled oscillator model reveals that the coupling strength is anisotropic and is significantly influenced by the spacer thickness between the graphene and BP nanoribbons. Thus, the heterostructures of BP and other 2D materials provide a new platform for the investigation of novel 2D plasmonic devices.

## 3.4. Other Layered Materials for Anisotropic and Hyperbolic Plasmons

Although BP is proposed to host hyperbolic plasmons in many theoretical works, many difficulties exist in the investigation of polaritons in real experiments, including the low carrier density, the large bandgap and the sensitivity to the air. The first two make the momentum of the polaritons in the hyperbolic regime extremely large, posing challenges for the detection by both the far and the near field methods. Although the carrier density can be increased by doping and gating, impurities and defects are usually introduced by these methods as well, which creates more damping channels for the plasmons. The last one makes BP easily degraded in the air, which increases the damping rate of plasmons. Therefore, other layered anisotropic materials with higher carrier concentrations, lower interband transition frequencies, and air-stability are needed to investigate the hyperbolic plasmons. Anisotropic plasmons were previously discussed in semiconducting 2D materials, such as BP. However, we can shift our attention to semimetals with in-plane anisotropic electronic structures. In contrast to the semiconducting BP, the semimetal films have higher carrier densities, which alleviates the difficulties in fabricating nanocavities with ultrasmall sizes. One category can be found in the layered topological semimetal materials with tilted Weyl or Dirac cones (**Figure 10**a), such as Type-II Weyl semimetals (e.g., $WTe_2$,[191-193] $T_d$-$MoTe_2$,[194, 195] $Mo_xW_{1-x}Te_2$,[196, 197] and $TaIrTe_4$[198]) and Type-II Dirac semimetals (e.g., $PtTe_2$,[199] and $PtSe_2$[200]). The magneto-transport,[201-207] optical measurements[208-211] and angle-resolved photoemission spectroscopy (ARPES) measurements[192-195, 197-200] reveal these Type-II semimetals with anisotropic electronic structures and simultaneous high mobility, which are essential for studying anisotropic 2D plasmons.

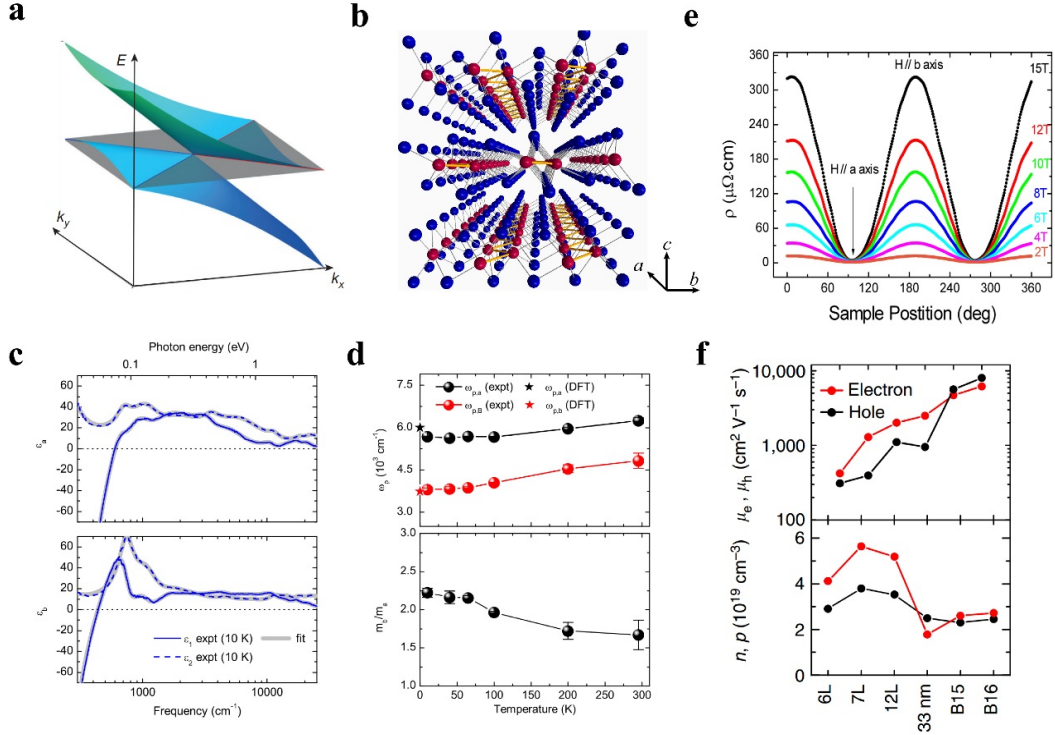

**Figure 10.** Anisotropic electronic structure in WTe$_2$. a) Illustration of the band structure of Type-II Weyl semimetal with highly tilted Weyl cones. Reproduced with permission.[191] Copyright 2016, Springer Nature. b) Schematic of the crystal structure of WTe$_2$. The W chains along the *a* axis are denoted by the yellow lines. c) The dielectric function along the *a* and *b* axes, extracted from the reflection data of single crystal WTe$_2$. The solid and dashed blue lines denote the real and imaginary parts. d) Temperature dependence of the bulk plasmon frequencies and the effective mass along the two principle axes, extracted from the reflection measurements. Reproduced with permission.[208] Copyright 2017, American Physical Society. e) The anisotropic magnetoresistance measured in WTe$_2$ flakes at 2 K. Reproduced with permission.[202] Copyright 2015, American Physical Society. f) Thickness dependence of the mobility and carrier density of WTe$_2$ at 250 mK. Reproduced with permission.[212] Copyright 2015, Springer Nature.

For example, the semimetal WTe$_2$[191-193] is composed of W chains running along the *a* axis, as shown in Figure 10b. The distance between adjacent W atoms along the *a* axis is much smaller than that along the *b* axis, leading to the anisotropic in-plane electronic structure. The optical properties in the *ab* plane are studied by the reflection measurements of single crystals.[208, 209] As shown in Figure 10c, the derived dielectric function is highly anisotropic along the two principle axes, which are composed of anisotropic intraband transitions (low energy components) and interband transitions. It is noted that the real parts of the dielectric functions along the two principle axes cross the zero line at different frequencies, indicating the possibility of natural in-plane hyperbolic regimes. In addition, the derived effective mass ratio and the bulk plasmon frequency (Figure 10d) are temperature dependent, indicating possible tuning of the plasmons by temperature. The anisotropic electron dynamics can also be manifested by the anisotropic magnetoresistance[201, 202] measured in WTe$_2$ single crystals (Figure 10e). Ultrahigh mobility was reported with a value of approximately 10,000 cm$^2$/(V·s) for the bulk at low temperatures (Figure 10f).[212] The high carrier mobility is essential for the observation and application of plasmons, since it determines the linewidth of the localized plasmon resonance mode and the travelling length of the propagating plasmon polariton.[104, 105]

For the semimetal phase of MoTe$_2$, Chen et al. reported an anisotropic magnetoresistance in the low temperature T$_d$ phase, with an electron mobility of approximately 5000 cm$^2$/(V·s).[203] The in-plane effective mass anisotropy ratio increases with a decreasing temperature, and reaches approximately 4 at low temperatures. In addition to the anisotropic intraband transitions (due to anisotropic effective mass), the anisotropic interband transitions in MoTe$_2$ are demonstrated by the photocurrent measurements. Lai et al. observed an anisotropic photoresponse in the semimetal phase of MoTe$_2$ via excitation in a broadband spectral range from 532 nm to 10.6 μm.[210] In addition, the anisotropy ratio of the photocurrent is excitation-frequency-dependent, with a larger ratio in the long excitation wavelength. Moreover, MoTe$_2$ possesses the anisotropic semimetal phase (T$_d$ and 1T') and the isotropic semiconducting phase (2H). Due to the low energy difference between them, the phase transition can be induced by laser heating,[213] electrical[214] and mechanical[215] tuning, which adds tunability to the anisotropy of the plasmons in MoTe$_2$. Similar anisotropic magnetoresistance[205] and photoresponse[211] are also reported for TaIrTe$_4$.

It is worth noting that the mass ratio is not the only factor used to determine the hyperbolic regimes. As discussed above, to realize the hyperbolic 2D plasmons, one needs the coupling between the intraband and interband transitions.[79] Thus, the most critical value for determining the hyperbolic regime is not the effective mass ratio, but the anisotropy of the intraband and interband couplings along the two principle axes. In layered anisotropic semiconductor materials, such as BP, BP-analog materials (e.g., SnSe,[216] SnS,[217] GeSe,[218] and GeS[219]), the 1T phase of TMDCs (e.g., ReSe$_2$[220]), and the trichalcogenides (e.g., TiS$_3$[221]), strong anisotropic in-plane electronic properties have been demonstrated via anisotropic optical absorptions, polarization-dependent PL and anisotropic conductivities, while the low carrier densities (low intensity of intraband transitions) and relatively large band gap (from the mid-infrared to the visible spectrum) are against forming strong coupling between the intraband and interband excitations unless the plasmon frequencies are tuned to the vicinity of interband excitations. In such cases, the wave vector for plasmons would be extremely large, which is impractical in experiments. In contrast, the interband transition energies are much reduced in semimetal materials. As shown in Figure 10c, the lowest interband transition in WTe$_2$ is approximately 1000 cm$^{-1}$, which is close to the far-IR spectral range. An anisotropic photoresponse can be still observed by an excitation with a 10.6 μm wavelength in the semimetal phase of MoTe$_2$.[210] Combined with the relatively large carrier density, it is possible for semimetals to have a strong coupling between the intraband and interband transitions in the far and mid-IR range with moderate plasmon wave vectors, which facilitates the observation of plasmons by both far-field and near-field methods.

## 4. Conclusion

As a typical anisotropic 2D material, BP exhibits great potential in terms of its optical (interband transitions) and plasmonic (intraband transitions) properties. The intrinsic in-plane anisotropy not only gives another degree of freedom to photonic and optoelectronic devices, but it also gives rise to some unique properties that are not available in isotropic materials (e.g., hyperbolicity). Among the enormous family of 2D materials, anisotropic 2D materials are only beginning to emerge and are attracting much less attention compared with isotropic materials, such as graphene and TMDCs. This review aimed to stimulate further research interest in anisotropic 2D materials, especially in the discovery of new materials and the development of heterostructures, as well as the design of new-concept devices based on anisotropic 2D materials.

## Acknowledgments

H.Y. is grateful for the financial support from the National Natural Science Foundation of China (Grant Nos. 11874009, 11734007), the National Key Research and Development Program of China (Grant Nos. 2016YFA0203900 and 2017YFA0303504), the Strategic Priority Research Program of Chinese Academy of Sciences (XDB30000000), and the Oriental Scholar Program from Shanghai Municipal Education Commission. G.Z. acknowledges the financial support from the National Natural Science Foundation of China (Grant: 11804398) and the Fundamental Research Funds for the Central Universities. C.W. is grateful to the financial support from the National Natural Science Foundation of China (Grant: 11704075) and the China Postdoctoral Science Foundation.

## Competing financial interests

The authors declare no competing financial interests.